\documentclass[conference]{IEEEtran}
\IEEEoverridecommandlockouts
\usepackage{cite}
\usepackage{graphicx}
\usepackage{amsmath,amssymb,amsfonts}
\usepackage{algorithmic}
\usepackage{graphicx}
\usepackage{textcomp}
\usepackage{xcolor}
\usepackage{tikz}
\usepackage[algo2e,ruled,linesnumbered,boxed,commentsnumbered]{algorithm2e}
\newcommand{\tabitem}{~~\llap{\textbullet}~~}
\usepackage{latexsym}
\usepackage{adjustbox}
\usepackage{enumitem}
\usepackage{dcolumn}
\usepackage{booktabs}
\usepackage{amsmath}
\usepackage{multirow}
\usepackage{graphicx}
\usepackage{amssymb}
\usepackage{pifont}
\usepackage{amsmath,array,graphicx}
\usepackage{kantlipsum}
\def\BibTeX{{\rm B\kern-.05em{\sc i\kern-.025em b}\kern-.08em
    T\kern-.1667em\lower.7ex\hbox{E}\kern-.125emX}}
\usepackage{url}

\usepackage{breakurl}
\usepackage[breaklinks]{hyperref}

\begin{document}

\title{Internet of Things for Current COVID-19 and Future Pandemics: An Exploratory Study}
\author{
    \IEEEauthorblockN{Mohammad Nasajpour\IEEEauthorrefmark{1}, Seyedamin Pouriyeh\IEEEauthorrefmark{1}, Reza M. Parizi\IEEEauthorrefmark{2}, Mohsen Dorodchi\IEEEauthorrefmark{3}, Maria Valero\IEEEauthorrefmark{1}, Hamid R. Arabnia\IEEEauthorrefmark{4}}
    \IEEEauthorblockA{\IEEEauthorrefmark{1} Department of Information Technology, Kennesaw State University, Marietta, GA, USA
    \\mnasajp1@students.kennesaw.edu, \{spouriye, mvalero2\}@kennesaw.edu}
    \IEEEauthorblockA{\IEEEauthorrefmark{2} Department of Software Engineering and Game Development, Kennesaw State University, Marietta, GA, USA   \\rparizi1@kennesaw.edu}

       \IEEEauthorblockA{\IEEEauthorrefmark{3}Department of Computer Science,  University of North Carolina at Charlotte, NC, USA
    \\  mohsen.dorodchi@uncc.edu}
   
   \IEEEauthorblockA{\IEEEauthorrefmark{4}Computer Science Department, University of Georgia, Athens, GA, USA
    \\hra@uga.edu}
 \thanks{ \scriptsize \noindent Corresponding author: S. Pouriyeh (email: spouriye@kennesaw.edu).}   
}

\maketitle
\begin{abstract}
In recent years, the Internet of Things (IoT) has drawn convincing research ground as a new research topic in a wide variety of academic and industrial disciplines, especially in healthcare. The IoT revolution is reshaping modern healthcare systems by incorporating technological, economic, and social prospects. It is evolving healthcare systems from conventional to more personalized healthcare systems through which patients can be diagnosed, treated, and monitored more easily. The current global challenge of the pandemic caused by the novel severe contagious respiratory syndrome \mbox{coronavirus 2} presents the greatest global public health crisis since the pandemic influenza outbreak of 1918. At the time this paper was written, the number of diagnosed \mbox{COVID-19} cases around the world had reached more than 31 million. 
Since the pandemic started, there has been a rapid effort in different research communities to exploit a wide variety of technologies to combat this worldwide threat, and IoT technology is one of the pioneers in this area.  In the context of \mbox{COVID-19}, IoT enabled /linked devices/applications are utilized to lower the possible spread of \mbox{COVID-19} to others by early diagnosis, monitoring patients, and practicing defined protocols after patient recovery. This paper surveys the role of IoT-based technologies in \mbox{COVID-19} and reviews the state-of-the-art architectures, platforms, applications, and industrial IoT-based solutions combating \mbox{COVID-19} in three main phases, including early diagnosis, quarantine time, and after recovery.

\end{abstract}



\section{Introduction}
\label{intro}
The term “Internet of Things” (IoT) was first coined in a presentation about implementing Radio-frequency identification (RFID) in the Protector and Gamble company by Kevin Ashton for supply chain management \cite{ashton2009internet}. IoT is an advanced technology that can link all smart objects together within a network with no human interactions \cite{ali2015internet}. More simply, any object that can be connected to the internet for further monitoring or transferring data has the opportunity to be an IoT device \cite{haddadpajouh2019survey}.

In recent years, IoT has drawn convincing research ground as a new research topic in a wide variety of academic and industrial disciplines, especially in healthcare. The IoT revolution is reshaping modern healthcare systems, incorporating technological, economic, and social prospects. It is evolving healthcare systems from conventional to more personalized healthcare systems through which patients can be diagnosed, treated, and monitored more easily.

IoT is increasingly becoming a vital technology in healthcare systems where it can deliver lower expenses, a better quality of services, and advanced user experiences \cite{da2018internet, islam2015internet, hu2013application, Qi2017AdvancedIO}. As a result of its wide capabilities including tracking, identification and authentication, and data collection, the exponential growth of IoT in healthcare is expected to rise from USD 72 billion in 2020 to USD 188 billion in 2025 \cite{IoHTMarketSize2020, ali2015internet}.

The current global challenge of the pandemic caused by the novel severe respiratory syndrome coronavirus 2 presents the greatest global public health crisis since the pandemic influenza outbreak of 1918\cite{Lovelace2020}. 
According to the last report of the World Health Organization (WHO), as of September 2020, the number of confirmed \mbox{COVID-19} cases passed 31 million people with an approximate huge death toll number of 960,000 people \cite{COVIDCases2020}. This disease has similar symptoms as the flu such as fever, cough, and fatigue, which are essential to recognize for early diagnosis\cite{cdcsymptoms2020}. The incubation period of \mbox{COVID-19} takes from 1 to 14 days. Surprisingly, a patient without any symptoms can possibly be a transmitter of the \mbox{COVID-19} virus to others. This is when quarantining such people is necessary \cite{CDCQuarantine2020}. Moreover, the recovery period of this disease varies and depends on the patient’s age, underlying condition, etc., but in general it can take between 6 to 41 days \cite{wang2020updated}. While this disease has a high potential to be spread easily in comparison with similar diseases within the coronavirus family, there are many ongoing efforts and much research to mitigate the spread of this virus. In this context, IoT technology has been shown to be a safe and efficient way of dealing with the \mbox{COVID-19} pandemic \cite{peeri2020sars, singh2020internet, ting2020digital}.

Our goal in this study is to determine the role of IoT-based technologies in \mbox{COVID-19} and review the state-of-the-art architectures, platforms, applications, and industrial IoT-based solutions combating \mbox{COVID-19} in three main phases, including \textit{early diagnosis}, \textit{quarantine time}, and \textit{after recovery}.

Early detection and diagnosis can lead to lesser infection and, as a result, better health services for infected patients \cite{to2020consistent}. Quarantining confirmed or suspected cases and enforcing lockdowns can also decrease the number of \mbox{COVID-19} infections by separating infected people from others. Tracking \mbox{COVID-19} patients after recovery will benefit the monitoring of returning symptoms and the potential infectivity of these recovered cases\cite{xing2020post}.

The remainder of the paper is organized as follows. Section \ref{sec:Role of IoT} covers the importance of IoT during \mbox{COVID-19}. Section \ref{sec:phase1} highlights IoT technologies along with their categories for the phase of ``Early Diagnosis." Similarly, Section \ref{sec:phase2} and \ref{sec:phase3} review IoT technologies in ``Quarantine Time" and ``After Recovery" phases respectively. 
Finally, we discuss, outline future work, and conclude in sections \ref{sec:Discussion} and \ref{sec:Conclusion} respectively.

\section{Important Role of IoT in \mbox{COVID-19}}
\label{sec:Role of IoT}
Since early 2020, the world has been struggling with the pandemic caused by the novel severe respiratory syndrome coronavirus 2  by striving to control the unprecedented spread of the virus and develop a vaccine\cite{zhang2020unprecedented}. As most efforts to find a treatment or control the spread of the \mbox{COVID-19} have not shown acceptable results so far, there is a high demand for global monitoring of patients with symptomatic and asymptomatic \mbox{COVID-19} infection.

In recent years, IoT technology has received significant attention in the healthcare domain where it plays an important role in different phases of various infectious diseases\cite{christaki2015new}. In the current pandemic, as the contingency of \mbox{COVID-19} is high, there is an essential need for patients to be connected with and monitored by their physicians proactively in different phases of \mbox{COVID-19}. In this study, we investigate the role of IoT technology in response to \mbox{COVID-19} in three main phases including \textit{early diagnosis}, \textit{quarantine time}, and \textit{after recovery}.

During the first phase of \mbox{COVID-19}, which is early diagnosis\cite{phelan2020novel}, there is an essential need for faster diagnosis due to the high rate of contagiousness of COVID-19 where even an asymptomatic patient can easily spread the virus to others. The sooner the patient is diagnosed, the better the spread of the virus can be controlled, and the patient can receive appropriate treatment. In fact, IoT devices can speed up the detection process by capturing information from patients. This can be implemented by capturing body temperatures using different devices, taking samples from suspicious cases, and so on. 

The second phase, called quarantine time \cite{nussbaumer2020quarantine}, is an important period of this disease after the patient has been diagnosed with \mbox{COVID-19}, and he or she should be isolated for the course of treatment. IoT devices  in this phase can monitor patients remotely\cite{rahman2020defending} with respect to their treatments and stay at home orders by the authorities. They can also clean areas without human interactions. Examples of these types are the implementation of tracking wearable bands, disinfecting devices, etc.
 
According to the Centers for Disease Control and Prevention (CDC)\cite{CdcRecovery2020}, most people with mild symptoms can recover while staying at home without getting treatments, but there is no guarantee those people will not be reinfected after recovery. Reinfection might happen with different symptoms of \mbox{COVID-19}\cite{reinfection2020}. Concerning these possible reinfections in the after recovery phase, the chances of returning symptoms and the potential infectivity can be high\cite{xing2020post}. To prevent that happening, social distancing should be implemented by deploying IoT devices including bands, crowd monitoring devices, etc. to track people to ensure the appropriate distance is maintained.
In short, IoT technology during the \mbox{COVID-19} pandemic has proven its usefulness in assisting patients, healthcare providers, and authorities. In this section, we briefly explain the various IoT devices and applications including wearables, drones, robots, IoT buttons, and smartphone applications that are mainly utilized in the forefront of combating \mbox{COVID-19}. Table \ref{tab:IoT Technologies} lists the specifications of these technologies regarding this pandemic.

\subsection{Wearables}
Wearable technologies can be defined as the combination of electronics with anything that is able to be worn \cite{hayes_2020}. The definition presented by Juniper Research \cite{wearableDefinition2020} describes them as app-enabled computing technologies that receive and process input while they are either worn or stick to the body such as bands, glasses, watches, etc. These smart wearables were designed for different purposes in various domains such as healthcare, fitness, lifestyle, and so on \cite{wright2014wearable, wearableDefinition2020, berglund2016survey}. Although the privacy of data is still a significant issue for expanding these devices, it is predicted that healthcare providers will spend \$20 billion annually until 2023 on wearable IoT devices to monitor more patients\cite{wearableStat2020}.
IoT wearable devices cover a wide range of different smart wearable tools such as Smart Themormeters\cite{chamberlain2020real, tamura2018current}, Smart Helmets\cite{mohammed2020novel}, Smart Glasses\cite{mohammed20202019}, IoT-Q-Band\cite{Vibhutesh2020}, EasyBand\cite{tripathy2020easyband}, and Proximity Trace\cite{NED2020}. \mbox{Table \ref{tab:wearables}} shows all wearable devices regarding their classification with examples.

\subsection{Drones}
Drones are simply aircraft that are flown without any or very little human operation by remote monitoring \cite{kardasz2016drones}. In 1849, during a war between Italy and Austria, the first drone, which was a balloon equipped with bombs, was used\cite{naughton2007remote}. The drone is also known as an unmanned aerial vehicle (UAV) that works with the help of sensors, GPS, and communication services. The implementation of IoT within drones, known as the Internet of Drone Things (IoDT), makes it possible for drones to do a variety of tasks such as searching, monitoring, delivering, etc \cite{droneDefinition2019, nayyar2020internet}. Smart drones can be operated by a smartphone and a controller with a minimum of time and energy, which makes them efficient in different fields such as agriculture, military, healthcare, etc. Different types of IoT-based drones, including Thermal Imaging Drone\cite{mohammed2020toward}, Disinfectant Drone\cite{shawdesign}, Medical Drone\cite{zema2016medrone}, Surveillance Drone\cite{ding2018amateur}, Announcement Drone\cite{marr2020robots}, and Multipurpose Drone\cite{multipurpose2020} are used in the healthcare domain and, in particular, in the fight against \mbox{COVID-19},  will be discussed in this paper. An illustration of these types of drones, along with their examples, can be found in Table \ref{tab:drones}.

\begin{table*}[ht]

 \centering
 \caption{IoT enabled/linked Technologies during \mbox{COVID-19} }
  \centering
  \label{tab:IoT Technologies}
  \begin{adjustbox}{width=\textwidth}
  
  \begin{tabular}{lllll}
    \toprule
    Technology  & Description & Pros & Cons  \\ 
    \midrule 
    Wearables & An app-enabled technology for receiving and processing & \tabitem Consistent monitoring & \tabitem Security and privacy of data  \\
    & data that is worn on or stick to the body  & \tabitem Improving the quality of patient's medicare & \tabitem Short battery life \\
    & & \tabitem Safer and more efficient hospitals &  \\
    & & \tabitem Lowering hospital visits &   \\
    [.5\normalbaselineskip]
    \midrule 
    Drones & An aircraft equipped with sensors and cameras, GPS  & \tabitem Perform variety of tasks such as (searching, monitoring, delivering) & \tabitem Security issue (large unstructured data)  \\
    & and communication systems which is flown with less & \tabitem Reach to hard-access locations & \tabitem Quality of Service \\
    &  or no human interactions & \tabitem Lower the workers' interactions such as maintaining & \tabitem Low connections \\
    [.5\normalbaselineskip]
    \midrule 
    Robots & A programmable machine which can handles complex & \tabitem Lowering interactions by remote diagnosis and treatments & \tabitem Bias and Privacy concerns  \\
    &  actions like a living creature & \tabitem Maintaining such as cleaning and disinfecting & \tabitem Reduce mental health problems\\
    [.5\normalbaselineskip]
    \midrule 
    Smartphone Applications & An application software  designed to do limited tasks  & \tabitem Monitoring and Tracking & \tabitem Collected data privacy and Security \\
    & within a mobile device & \tabitem Cost-effective &  \\
    [.5\normalbaselineskip]
    \bottomrule
  \end{tabular}
  \end{adjustbox}

\end{table*}

\subsection{Robots}
According to the Merriam Webster dictionary\cite{robotDefinition2020}, a robot is defined as ``a machine that resembles a living creature in being capable of moving independently." As an advancement during the emergence of networked robots within the cloud, the Internet of Robot Things was implemented where they can do many different tasks to make life easier \cite{ray2016internet}. Regarding the current pandemic, robots can be categorized as Autonomous robots\cite{Autonomous2020covid}, Telerobots\cite{tavakoli2020robotics}, Collaborative robots\cite{cobot2020covid}, and Social robots\cite{SocialBot2020}. Table \ref{tab:robots} covers the fundamental aspects of these robots with examples.

\subsection{IoT Buttons}
This type of IoT device is a small, programmable button connected to the cloud through wireless communication\cite{chai2018internet}. Based on its written code on the cloud, this device can perform different repetitive tasks by pressing only one button. For example, one type of IoT buttons enable patients to complain if any hospital restrooms need cleaning by pressing a button only \cite{AWSIoTButton2020, chai2019internet}. Table \ref{tab:IoTbuttons} illustrates two implementations of these button during \mbox{COVID-19} phases.

\subsection{Smartphone Applications}
Smartphone applications are application software designed to do limited tasks within a mobile device such as a smartphone \cite{mobileAppDef2018,8875296}. Since there are 3.5 billion active smartphones in 2020, these IoT-based smartphone applications could be very efficient in various domains such as healthcare, retail, agriculture, etc.  \cite{turner2020many, el2017smartphone, smartphoneApp2019,parizi2018cyberpdf}. Many smartphone applications have been developed for healthcare domain, and some of them have been used in response to \mbox{COVID-19}, as illustrated in Table \ref{tab:smartphoneApp}, namely nCapp\cite{bai2020chinese}, DetectaChem\cite{DetectaChem2020}, Stop Corona\cite{Stopcorona2020}, Social Monitoring\cite{Russia2020}, Selfie app\cite{SelfieApp2020}, Civitas\cite{Civitas2020}, StayHomeSafe\cite{HongKong2020}, AarogyaSetu\cite{AarogyaSetu2020}, TraceTogether\cite{TraceTogether2020}, Hamagen\cite{Israel2020}, Coalition\cite{coalition2020}, BeAware Bahrain\cite{Bahrain2020}, eRouska\cite{erouska2020}, and Whatsapp\cite{WhatsApp2020}.


\section{Phase I: Early Diagnosis} \label{sec:phase1}
The key to combating \mbox{COVID-19} is to diagnose it early to prevent spreading the virus widely. This will substantially help healthcare providers to arrange better treatment plans, save more lives, and reduce contamination and infections  \cite{sabeti2020early}.
The first step in the early diagnosis of \mbox{COVID-19} is understanding its symptoms. According to the CDC \cite{cdcsymptoms2020}, as of September 2020, \mbox{COVID-19} has a wide range of symptoms including fever or chills, cough, shortness of breath or difficulty breathing, fatigue, muscle or body aches, headache, the new loss of taste or smell, sore throat, congestion or runny nose, nausea or vomiting, and diarrhea. Among them, fever or high body temperature is one of the most common symptoms of \mbox{COVID-19} when the measured temperature exceeds 38 Degrees Celsius or 100.4 Degrees Fahrenheit \cite{cdcSympDefinitions2020}.
IoT devices can make the detection process faster and more efficient by capturing data within their sensors and then analyzing the data for patients, healthcare providers, and authorities to diagnose, control, and ultimately stop this contagious disease \cite{IoTDetection2020}. Different IoT devices can be used to capture some of the aforementioned symptoms at an early stage, which will be discussed in the next subsections.

\subsection{Wearables}
Using wearable devices is considered as an efficient way in response to the need for early diagnosis during this pandemic \cite{guk2019evolution}. Developing these devices has had a remarkable impact on the early detection of diseases. For example, a wearable IoT device can confirm whether respiratory signs of a patient is normal or not. With this knowledge, the patient can notice any changes in his/her health situation and then decide to make a medical appointment before any other symptoms appear\cite{haghi2017wearable}. In fact,  \mbox{COVID-19} pandemic might be easier to fight  using appropriate wearable devices.

\subsubsection{Smart Thermometers}
A wide range of IoT smart thermometers has been developed to record constant measurements of body temperatures. These low-cost, accurate, easy to use devices could be worn or stick to the skin under clothing \cite{tamura2018current}. They are usually offered in different forms such as touch, patch, and radiometric \cite{tamura2018current}. The use of these devices can be extremely helpful in the early detection of suspicious cases. Also, since the use of infrared thermometers for capturing body temperature can possibly spread the virus more due to the closeness of patients and health care providers, using smart thermometers is highly recommended \cite{mohammed2020toward}.

\begin{figure}[h]
\centering
    \includegraphics[scale=.15]{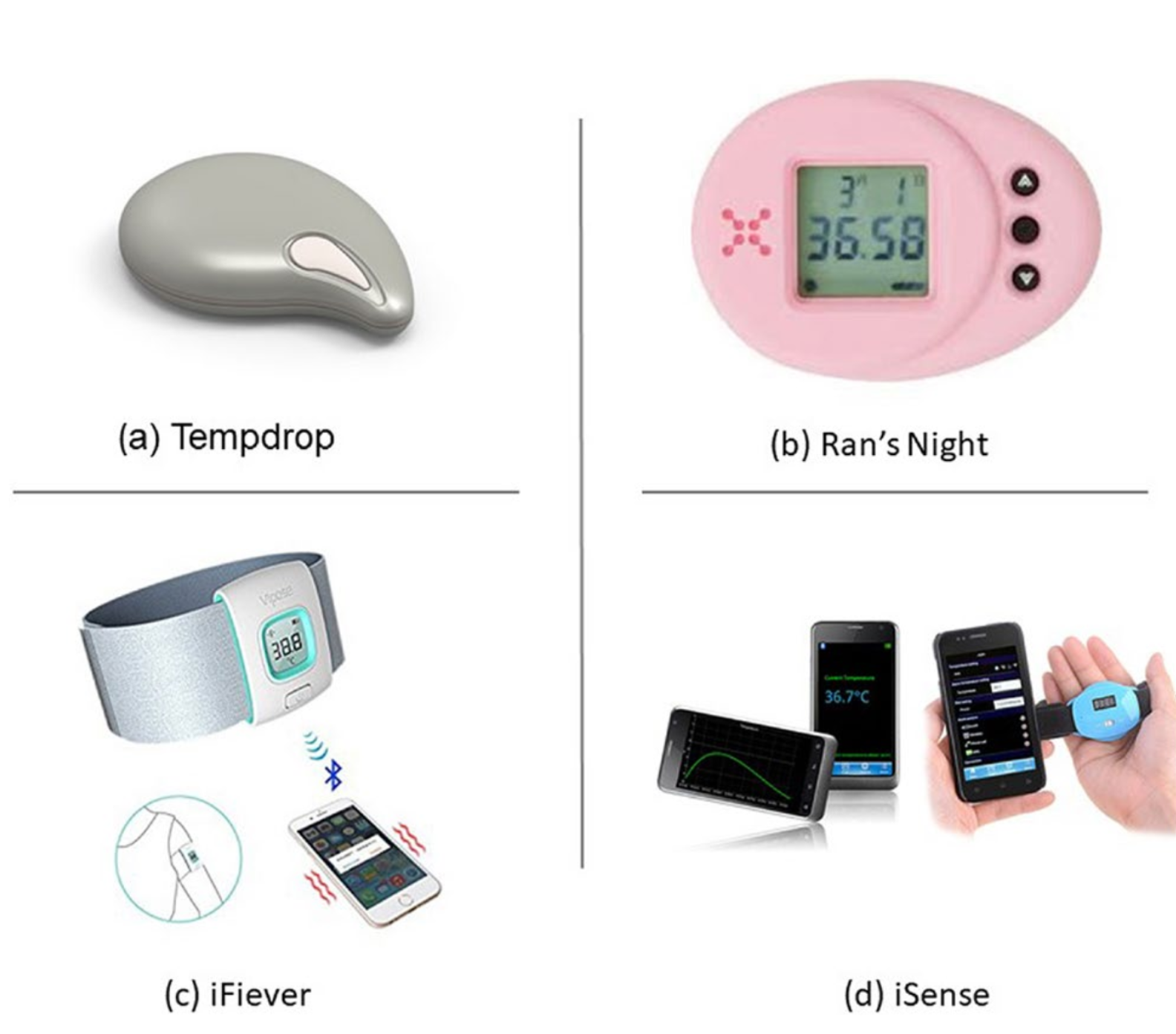}
    \caption{Wearable Smart Thermometers\cite{tamura2018current}.}
    \label{image:WearableSmart}
\end{figure}

According to \cite{UShealthweather2020}, Kinsa's thermometers have been widely used in homes, and the producer is now able to predict the most suspicious areas (contaminated with \mbox{COVID-19}) in each state of the USA based on the recorded temperature of people\cite{mcneil2020can, Kinsa, Gold2020, chamberlain2020real}. Other smart thermometers such as Tempdrop, Ran's Night, iFever, and iSense (shown in Fig. \ref{image:WearableSmart}) are able to report body temperature at any time on a smartphone. Using these devices in people's daily lives can improve the chance of diagnosing new patients at early stage.

\subsubsection{Smart Helmet}
\begin{figure} \centering
    \includegraphics[scale=.25]{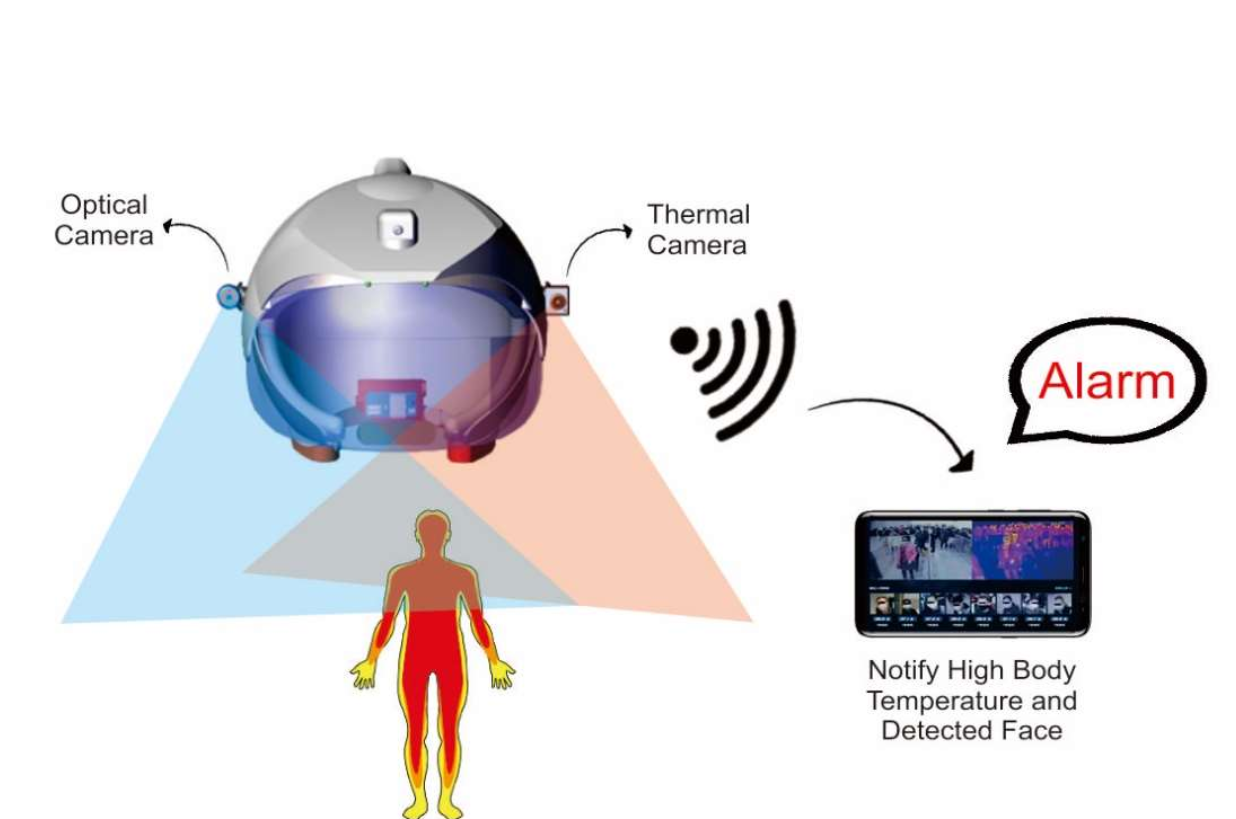}
    \caption{Smart Helmet captures temperatures using thermal and optical camera\cite{mohammed2020novel}.}
    \label{image:smartHelmet}
\end{figure}

During \mbox{COVID-19} pandemic, using  wearable smart helmets with a thermal camera has shown to be safer compared to an infrared thermometer gun due to lower human interactions \cite{mohammed2020toward}. In this device, when the high temperature is detected by the thermal camera on the smart helmet, the location and the image of the person's face are taken by an optical camera. Then, they are sent to the assigned mobile device with an alarm as shown in Fig. \ref{image:smartHelmet}, so that the health officer can distinguish the infected person, and authorities can take action \cite{mohammed2020novel}. Additionally, Google Location History can be incorporated with the smart helmet to find the places visited by the suspected person after detection \cite{ruktanonchai2018using}. Countries such as China, UAE, and Italy have implemented this wearable device to monitor crowds within two meters from passers-by\cite{helmetUse2020}. Interestingly, this model has shown good results. For example, KC N901 is a smart helmet produced in China that has an accuracy of 96 percent for high body temperature detection \cite{helmetUse2020}.

\begin{table*}[ht]
 \centering
 \caption{IoT enabled/linked Wearable Devices during \mbox{COVID-19} } 
  \centering
  \label{tab:wearables}
 \begin{adjustbox}{width=\textwidth}
  \begin{tabular}{llllc}
    \toprule
    Model & Type  & Capability & Examples & Phase \\ 
    \midrule 
    \cite{mohammed2020toward, Kinsa, chamberlain2020real, tamura2018current} & Smart Thermometers & \tabitem Temperature monitoring  & Kinsa, Tempdrop, Ran’s Night  &  I \\
    & & \tabitem Increasing the diagnosis rate & iFever, iSense\\ [.5\normalbaselineskip]
    \midrule
    \cite{mohammed2020novel, helmetUse2020} & Smart Helmet & \tabitem Temperature monitoring  & KC N901 in China & I \\
    & & \tabitem Capturing location and face image &  \\
    & & \tabitem Less human interactions &  \\[.5\normalbaselineskip]
    \midrule
    \cite{mohammed20202019, bright2020chinese}& Smart Glasses & \tabitem Temperature monitoring and capturing  & Rokid in China &  I \\
    & & \tabitem Less human interactions & Vuzix \& Onsight \\ [.5\normalbaselineskip]
    \midrule
    \cite{Vibhutesh2020, HongKong2020} & IoT-Q-Band & \tabitem Tracking quarantined objects in case of absconding  & Hong Kong electronic wristband &  II  \\ 
    & & \tabitem Cost-effective tracking & Electronic ankle bracelet in USA\\
    & & \tabitem Destructible & 
    \\[.5\normalbaselineskip]
    \midrule
    \cite{tripathy2020easyband, PactWristband2020} & EasyBand & \tabitem Monitoring social distancing practice by people  & Pact wristband  &  III \\ 
    & & \tabitem Alert the danger of closeness by LED &  \\[.5\normalbaselineskip]
    \midrule
    \cite{NED2020, ProximityExample2020, Proximity22020, instantTrace2020} & Proximity Trace & \tabitem Monitoring workers for the social distancing practice  & Hardhat TraceTag &  III \\
    & & \tabitem Tracing contacts of contaminated employee & Instant Trace & \\ [.5\normalbaselineskip]
  
    \bottomrule
  \end{tabular}
  \end{adjustbox}

\end{table*}
\subsubsection{Smart Glasses}
\begin{figure}[h] \centering
    \includegraphics[scale=0.10]{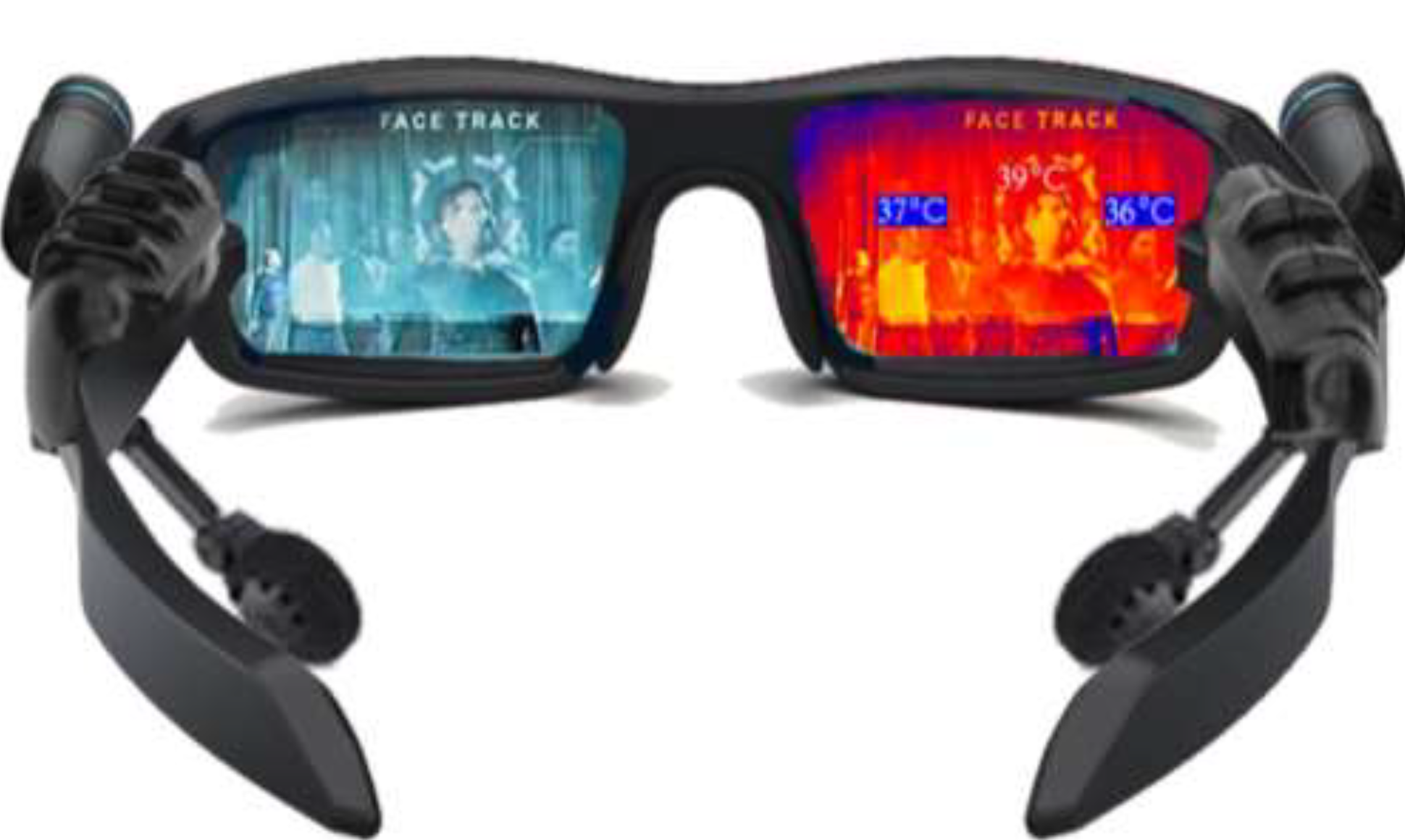}
    \caption{Smart Glasses Temperature Capturing\cite{mohammed20202019}.}
    \label{image:Smartglass}
\end{figure}

\begin{figure} \centering
    \includegraphics[scale=1.10]{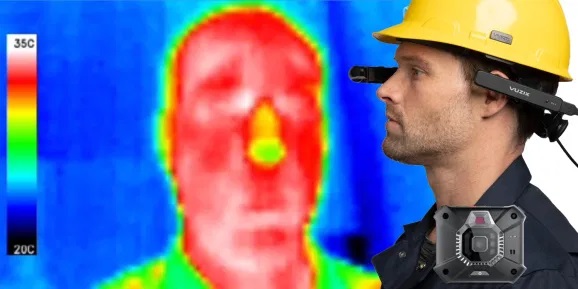}
    \caption{Vuzix Smart Glass\cite{horwitz2020vuzix}.}
    \label{image:Vuzixglass}
\end{figure}

Another type of wearable device is the IoT-based smart glasses as shown in Fig. \ref{image:Smartglass}. In comparison with thermometer guns, smart glasses have lesser human interactions. Optical and thermal cameras have been used in smart glasses to monitor crowds\cite{mohammed20202019} and the inbuilt face detection technology makes tracking procedure easier after detecting suspicious cases. In fact, this allows detecting the identification of the suspicious case (person with high temperature). Additionally, Google Location History can empower further actions with more reliability by capturing the places visited by the suspicious case \cite{mohammed20202019}.
Among different smart glasses, Rokid\cite{bright2020chinese}, smart glasses with infrared sensors, have the ability to monitor up to 200 people. Another example of this device is the combination of Vuzix smart glasses with the Onsight Cube thermal camera (see Fig.\ref{image:Vuzixglass}). These devices work together to monitor crowds to detect people with high temperatures and provide their information to medical centers or authorities \cite{Vuzix2020}.

\subsection{Drones}
\begin{figure}[h] \centering
    \includegraphics[scale=.22]{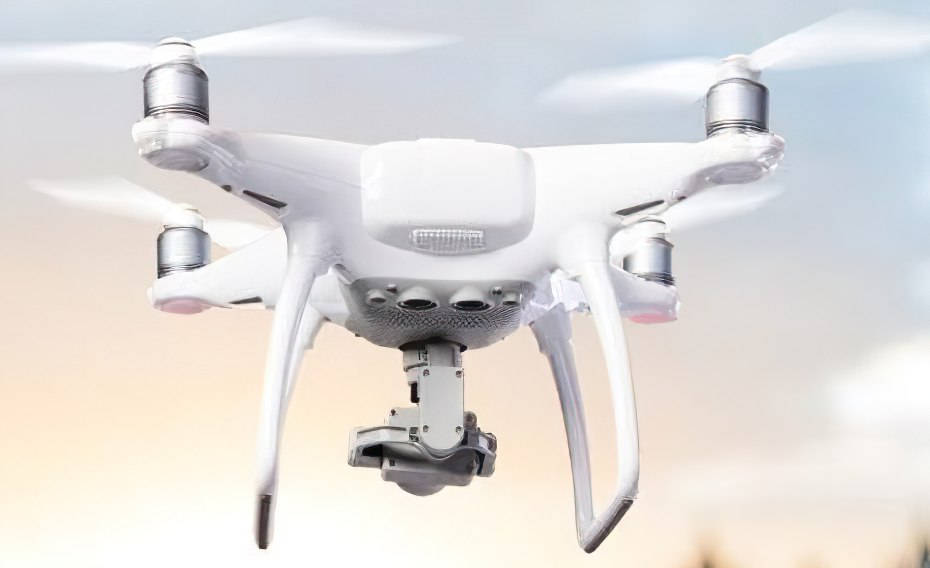}
    \caption{Thermal Imaging Drone\cite{Pandemic2Drone2020}.}
    \label{image:Pandemicdrone}
\end{figure}

In general, finding infected people in a crowd is important in early diagnosis and control of \mbox{COVID-19} \cite{haleem2020areas}. Using Unmanned Ariel Vehicles (UAV) and, especially, IoT-based drones is another common way to speed up the process of finding contaminated people and zones during this pandemic.  Drone technology can reduce human interactions and can reach hard-to-access locations \cite{chamola2020comprehensive}.
The Thermal Imaging Drone as shown in Fig. \ref{image:Pandemicdrone} was designed for capturing the temperature of people in crowds and can be used in the early diagnosis phase. This type of drone can be combined with Virtual Reality as a wearable device to identify people with high temperatures(fevers). 
This device not only reduces human interactions, but it also uses less time compared to thermometer gun devices\cite{mohammed2020toward}. One example of this device is the Pandemic Drone application developed by a Canadian company\cite{PandemicDroneCompany} for remote monitoring and detecting any cases of infection by capturing temperature, respiratory signs such as heart rate, and any sneezing or coughing \cite{cozzens2020pandemic, Pandemic2Drone2020}.



\begin{table*}[h]

 \centering
 \caption{IoT enabled/linked Drone Devices during \mbox{COVID-19}}
  \centering
  \label{tab:drones}
  \begin{adjustbox}{width=\textwidth}
  \begin{tabular}{llllc}
    \toprule
    Model & Type  & Capability & Examples & Phase  \\ 
    \midrule 
    \cite{mohammed2020toward, 2Drones2020} & Thermal Imaging Drone & \tabitem Temperature capturing in the crowd  & Pandemic Drone &  I  \\
    & & \tabitem Less human interactions & \\ [.5\normalbaselineskip]
    \midrule
    \cite{disinfect2020, 2Drones2020} & Disinfectant Drone & \tabitem Sterilizing contaminated areas  & DJI & II  \\
    & & \tabitem Preventing health workers from being infected &  \\
    & & \tabitem Less human interactions &  \\[.5\normalbaselineskip]
    \midrule
    \cite{chinadrone2020, medicalDrone2020, DeliveryDrone2020} & Medical/Delivery Drone & \tabitem  Reducing the hospital  visits & Delivery Drone Canada & II, III \\
    & & \tabitem Increasing accessibility to treatments &  \\ [.5\normalbaselineskip]
    \midrule
    \cite{marr2020robots, Cyient2020} & Surveillance Drone & \tabitem Crowd social distancing monitoring & MicroMultiCopter  & III  \\ 
    & & & Cyient & 
    \\[.5\normalbaselineskip]
    \midrule
    \cite{1Drones2020, 2Drones2020} & Announcement Drone & \tabitem Broadcasting information about \mbox{COVID-19} & Broadcasting drone in Spain and Kuwait  & III \\ 
    [.5\normalbaselineskip]
    \midrule
    \cite{multipurpose2020} & Multipurpose Drone & \tabitem Temperature capturing  & Corona Combat &  I, II, III \\
    & & \tabitem Disinfecting areas & & \\
    & & \tabitem Crowd monitoring & &  \\
    & & \tabitem Broadcasting information &  & \\
    [.5\normalbaselineskip]
  
    \bottomrule
  \end{tabular}
  \end{adjustbox}
   
\end{table*}

\subsection{Robots}
\begin{figure}[h] \centering
    \includegraphics[scale=.15]{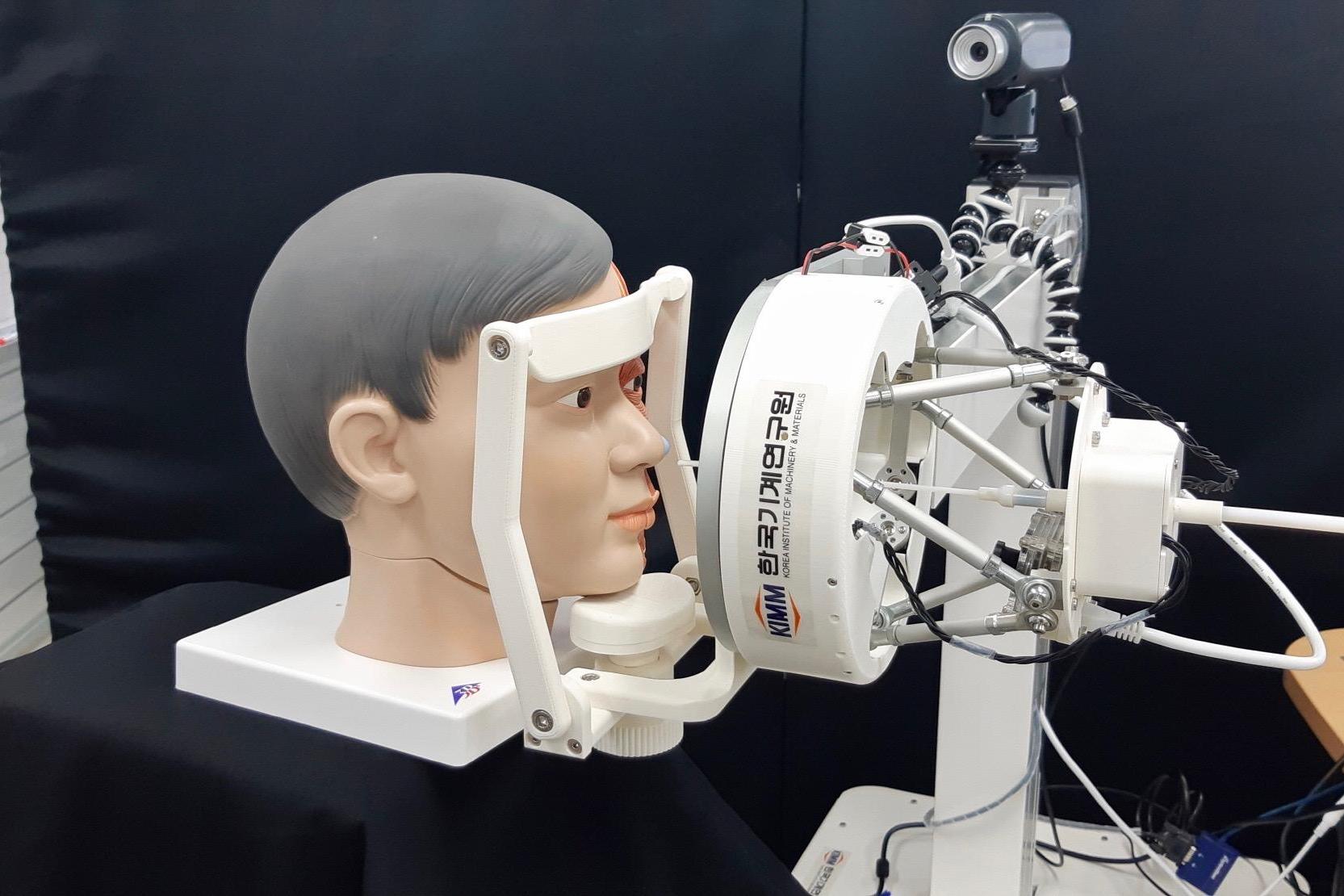}
    \caption{Autonomous Swab Test Robots\cite{Autonomous2020swab}.}
    \label{image:robotSwab}
\end{figure}

Using robots linked to IoT to assist early diagnosis is a remarkable use of these devices because they can help health workers by processing patients' treatments and lowering work stress levels \cite{yang2020combating}. Without the interaction of humans, the autonomous robot can help fight in all \mbox{COVID-19} phases. In the first phase, it can help the process of diagnosis by collecting throat swabs samples from patients with the advantage of preventing medical staff at risk (close contact with patients) \cite{Autonomous2020covid}. Fig. \ref{image:robotSwab} depicts how this process works. An example of this device, the Intelligent Care Robot, has been developed through a partnership between two companies, Vayyar Imaging\cite{vayyar2020} and Meditemi\cite{meditemi2020}. This device detects symptoms of COVID-19 in 10 seconds by using touchless quick scanning of a person within a distance of 1 meter to capture respiratory signs and temperature \cite{IntelligentCareRobot2020}.

\begin{table*}[]

 \centering
 \caption{IoT enabled/linked Robot Devices during \mbox{COVID-19}}
  \centering
  \label{tab:robots}
  \begin{adjustbox}{width=\textwidth}
  \begin{tabular}{llllc}
    \toprule
    Model & Type & Capability & Examples & Phase  \\ 
    \midrule 
    \cite{tavakoli2020robotics, ackerman2020autonomous, Kim2020, Danish2020, Autonomous2020covid} & Autonomous Robots & \tabitem Detecting symptoms & Intelligent  Care  Robot &  I, II, III  \\
        & & \tabitem Controlling social distancing & Spot Robot\\
    & & \tabitem Preventing medical staff being infected &  \\ 
    & & \tabitem Disinfecting and sterilizing contaminated areas in hospitals & \\
    & & \tabitem Bringing patients’ treatments & \\ 
    & & \tabitem Checking patients' respiratory signs & \\
    & & \tabitem Collecting swab tests & \\
    [.5\normalbaselineskip]
    \midrule
    \cite{tavakoli2020robotics, yang2020keep} & Telerobots & \tabitem Reducing the risk of infection for medical staff  & DaVinci surgical robots & II  \\
    [.5\normalbaselineskip]
    \midrule
    \cite{tavakoli2020robotics, cobot2020covid, EconomicTimes2020} & Collaborative Robots & \tabitem  Lower healthcare workers' fatigue & Asimov Robotics & II \\
    & & \tabitem Disinfecting hard-to-reach areas & eXtremeDisinfection Robot \\ [.5\normalbaselineskip]
    \midrule
    \cite{tavakoli2020robotics, SocialBot2020, Paro2020} & Social Robot & \tabitem Reducing mental strain & Paro  &  II \\ 
    [.5\normalbaselineskip]
  
    \bottomrule
  \end{tabular}
  \end{adjustbox}
   
\end{table*}
\subsection{IoT Buttons}
\begin{figure} \centering
\centering
    \includegraphics[scale=.25]{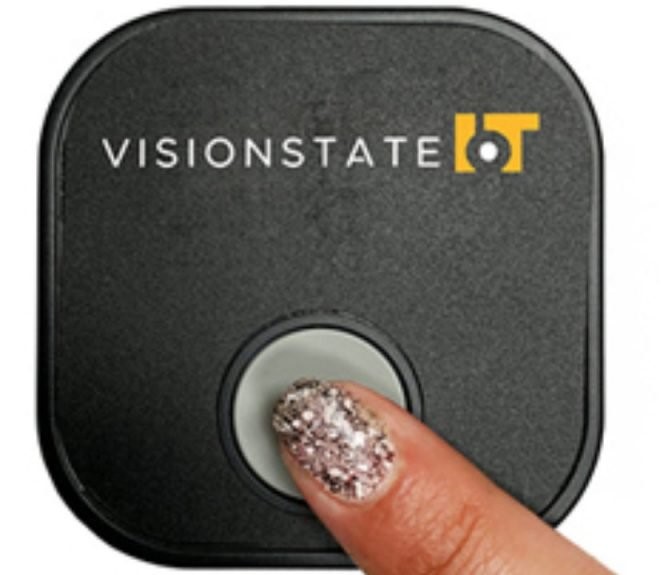}
    \caption{Wanda QuickTouch IoT Button\cite{Mahashreveta2020}.}
    \label{image:IoTButton}
\end{figure}

IoT Button, in general, is a programmable device that can be used for repetitive tasks. During this pandemic, IoT buttons can play an important role in alerting the authorities or family of a patient about any contaminated area or any emergency. For example, an IoT button, produced by Visionstate\cite{VisionState2020}, called Wanda QuickTouch (Fig. \ref{image:IoTButton}), were deployed as a cleaning alert system in hospitals. They are designed for alerting authorities in case of any concerns related to essential sanitation or public safety. 

\subsection{Smartphone Applications}
Smartphone applications enabled with IoT using  information such as Global Positioning System(GPS) and Geographic Information System(GIS), etc. for tracking purposes have been widely used during the COVID-19 pandemic in order to increase the chance of detecting infected people \cite{el2017smartphone}.
Implementing smartphone applications using the Internet of Medical Things (IoMT) will assist patients by providing them proper treatments while they are home. Additionally, it enables health care workers and authorities to monitor patients and the spread of disease more easily. People can upload their health information to the cloud adopted by IoT and get health advice from hospitals online. Using this platform, patients can be cured at home without expanding the contamination. It costs less than having a physical appointment at hospitals and allows the governments to take better action to manage the pandemic in the future\cite{yang2020combining}. Since the start of the pandemic, some smartphone applications have been developed for \mbox{COVID-19} diagnosis and monitoring that will be discussed in the coming sections. 

\begin{table*}[ht]

 \centering
 \caption{IoT Buttons during \mbox{COVID-19} }
  \centering
  \label{tab:IoTbuttons}
  \begin{adjustbox}{}
  \begin{tabular}{llllc}
    \toprule
    Model & Type  & Capability & Example & Phase  \\ 
    \midrule 
     \cite{Mahashreveta2020, Julian2020, Anasia2020, VisionState2020} & (1) & \tabitem Alerting the authorities or family & Wanda QuickTouch & I  \\
    [.5\normalbaselineskip]
    \midrule
    \cite{SefucyIoTButton2020} & (2) & \tabitem Alerting healthcare provider in case of an emergency  & Sefucy & II  \\
    [.5\normalbaselineskip]
  
    \bottomrule
  \end{tabular}
  \end{adjustbox}
   
\end{table*}

\subsubsection{nCapp}
\mbox{COVID-19} Intelligent Diagnosis and Treatment Assistant Program (nCapp) was developed in China using  Internet of Medical Things on a cloud platform. This cellphone application is an automated diagnosis system with eight functions that can be selected by the user. nCapp can automatically generate a diagnosis report based on requested data and questionnaires submitted by patients. 
Diagnosis is categorized into three cases: confirmed, suspected, or suspicious. For the confirmed cases, there are four conditions, including ``mild, moderate, severe, and critical,” which are determined by a physician. Special treatments for these conditions and other types of cases are defined as well. Other positive points of this program include updating its own database in order to improve its diagnosis, making consultation possible for all health workers, making sure all patients are safe in the long term, and finally, having all these abilities publicly available. In general, by using nCapp the diagnosis can be done faster and the spread of disease can be controlled easier \cite{bai2020chinese}.

\subsubsection{MobileDetect}
The high demand for a system that can identify infected people has led to the implementation of MobileDetect app \cite{DetectaChem2020}. MobileDetect which is compatible with a wide variety of smartphones is designed to detect and control the spread of COVID-19. Using this application, users can easily take the test at home utilizing a nasal swab. The results of the test will show up on the smartphone application within 10-30 minutes determining the user's health situation regarding \mbox{COVID-19}. Then, the user is able to send the results with any additional information needed to his/her physician or healthcare professional for further action. This smartphone testing kit authorized by the Food and Drug Administration (FDA) under emergency access can be helpful during the first phase of a pandemic by lowering the spread of the virus \cite{DetectaChem2020, fda2020fact}.

\subsubsection{Stop Corona}
Besides all the implementations for early case detection, another approach is having a database of captured daily health reports. The reports include contact with others, symptoms, and locations. The Stop Corona application\cite{Stopcorona2020} predictive heatmaps based on the disease spots. This application collects information from its users about their daily health status and generate report and heatmaps based on that. The generated report will be accessible only to health authorities. Consequently, once a user shows a new symptom and announces it, the case will be appear on the new report and ultimately authorities will be able to take proper action and detect the contaminated area faster due to the reported new symptoms.


\begin{table*}[ht]

 \centering
 \caption{IoT enabled/linked smartphone applications during \mbox{COVID-19} }
  \centering
  \label{tab:smartphoneApp}
  \begin{adjustbox}{width=\textwidth}
  \begin{tabular}{lllcl}
    \toprule
    Model & Application & Function & Origin & Phase \\ 
    \midrule 
    \cite{bai2020chinese} & nCapp & \tabitem Keeps database updated & China & I  \\
    & & \tabitem Provide available consulting \\
    & & \tabitem Controlling patient’s health in long-term &  \\ 
    [.5\normalbaselineskip]
    \midrule
    \cite{DetectaChem2020} & DetectaChem & \tabitem Taking \mbox{COVID-19} low-cost tests within a kit joined with a smartphone application & USA & I  \\
    [.5\normalbaselineskip]
    \midrule
    \cite{Stopcorona2020} & Stop Corona & \tabitem Getting daily health reports including contact, symptoms and location & Croatia & I  \\
    & & \tabitem Building a map with high risk spots \\
    [.5\normalbaselineskip]
   \midrule
    \cite{Russia2020, RussiaPrivacy2020, Meduza2020, kupper2011geofencing} & Social Monitoring & \tabitem Track patients with the diagnose of \mbox{COVID-19} & Russia & II \\
    & & \tabitem Access to the user's information by government (privacy concern) \\
    [.5\normalbaselineskip]
    \midrule
    \cite{SelfieApp2020, geolocation2020, bruce1986understanding} & Selfie app & \tabitem Monitoring patients by asking randomly to send selfies & Poland & II  \\
    [.5\normalbaselineskip]
    \midrule
    \cite{Civitas2020, 2Civitas2020} & Civitas & \tabitem Determining perfect time for suspected cases to leave for essentials & Canada & II  \\
    [.5\normalbaselineskip]
    \midrule
    \cite{HongKong2020, kupper2011geofencing} & StayHomeSafe & \tabitem  Monitoring arrivals at the airport with use of smartphone application and a wristband & Hong Kong & II  \\
    [.5\normalbaselineskip]
    \midrule
    \cite{AarogyaSetu2020, singh2020internet, 1Aarogya2020, 2Aarogya2020} & Aarogya Setu & \tabitem Linking people and health services better & India & III  \\
    [.5\normalbaselineskip]
    \midrule
    \cite{TraceTogether2020,Trace2020} & TraceTogether & \tabitem Capturing the people who were close to the user with encrypted IDs & Singapore & III  \\
    & & \tabitem Accessing to the user's information by government (privacy concern) \\
    & & \tabitem Notifying people who were in close contact with user if user is infected \\
    [.5\normalbaselineskip]
    \midrule
    \cite{Israel2020} & Hamagen & \tabitem  Finding out if the user has been in a close contact with a positive tested for \mbox{COVID-19} & Israel & III   \\
    [.5\normalbaselineskip]
    \midrule
    \cite{coalition2020} & Coaltion & \tabitem Securely notifying about detected cases who users have been in contact with & USA & III   \\
    [.5\normalbaselineskip]
    \midrule
    \cite{Bahrain2020, Bahrain2020second} & BeAware Bahrain & \tabitem Alerting people with close contact of infected person & Bahrain & III   \\
    & & \tabitem Track the self-isolated people \\
    & & \tabitem Location services must be ON \\
    [.5\normalbaselineskip]
    \midrule
    \cite{erouska2020, Erouska2020github} & eRouska(smart quarentine) & \tabitem Capturing physical contacts between user and people & Czech Republic & III   \\
    [.5\normalbaselineskip]
    \midrule
    \cite{WhatsApp2020} & Social Media - Whatsapp & \tabitem Provide healthcare support without visiting hospital & Singapore & I, II, III  \\
    & & \tabitem Available consulting with the physician \\
    [.5\normalbaselineskip]
    \bottomrule
  \end{tabular}
  \end{adjustbox}
   
\end{table*}

\section{Phase II: Quarantine Time} \label{sec:phase2}
 
After the process of detection, it is necessary to isolate and then monitor the patients either in a hospital or at home. The quarantine does not only apply to confirmed cases but also can be considered for suspected patients and even different areas or cities or countries \cite{CDCQuarantine2020}. This is done to prevent possible transmission from suspected cases (asymptomatic cases) or areas to others. Using IoT devices in this phase could help mitigate serious challenges such as spreading the virus by monitor patients efficiently and control their respiratory signs, heart rate, blood pressure, and so on \cite{wilder2020isolation, quarantine2020, vaishya2020artificial}.

\subsection{Wearables}
\begin{figure}[h] \centering 
    \includegraphics[scale=.26]{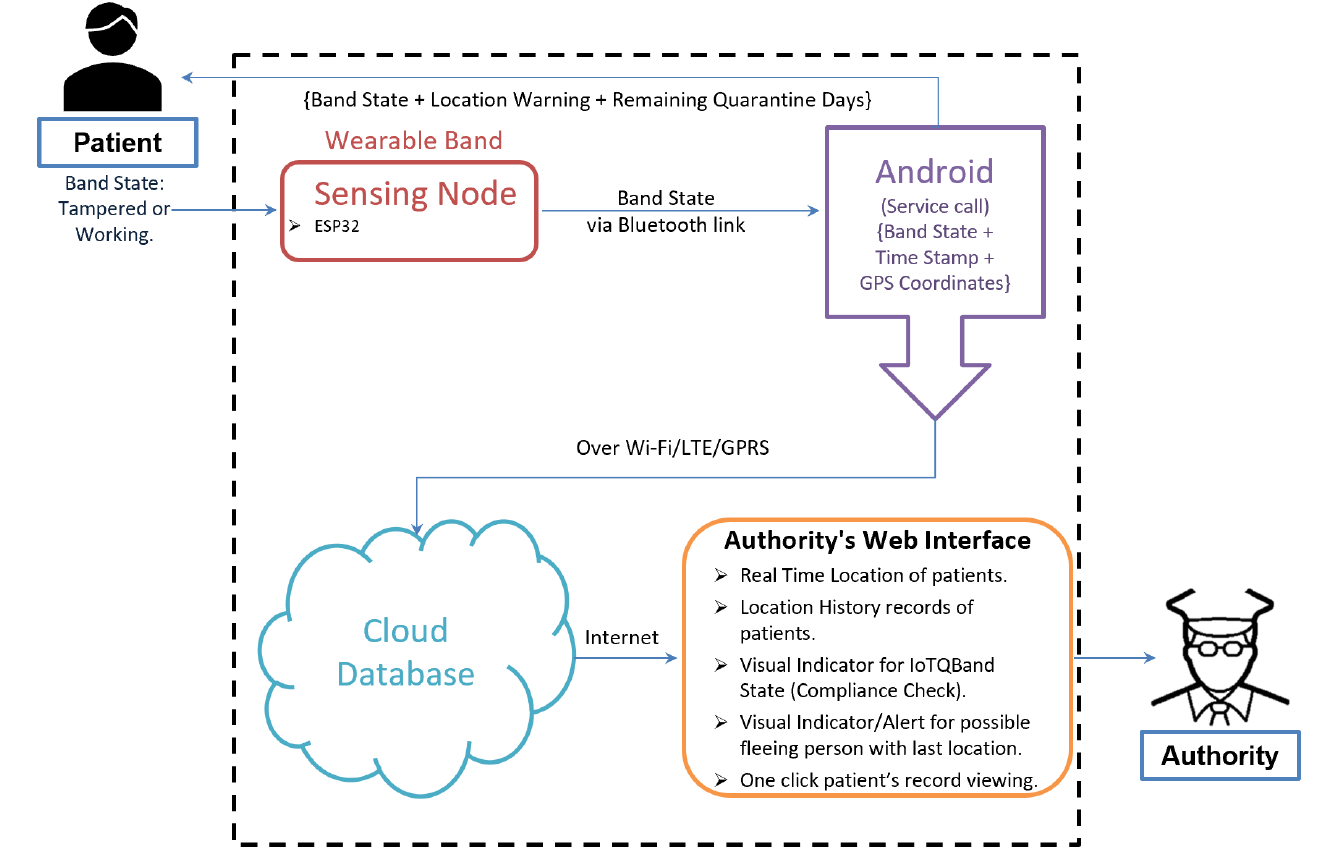}
    \caption{IoT-Q-Band workplace classification\cite{Vibhutesh2020}.}
    \label{image:IoT-Q-Band}
\end{figure}

Quarantine time for confirmed or suspected cases is vital because there is a chance of spreading the virus to other people by those cases \cite{nussbaumer2020quarantine}. IoT wearable bands have shown promising results to prevent patients from leaving the quarantine areas. 
Using wearable bands is a cost-effective solution for tracking cases. This device is connected to patience’s smartphone application through Bluetooth during the quarantine period and health care authorities can usually monitor all cases every two minutes using a web interface. Additionally, if a patient does not have the band on his or her arm or leg, or maybe leaves the quarantine area, an alert will be sent to notify the authorities and they have permission to call the patient for clarification of the situation. Fig. \ref{image:IoT-Q-Band} shows a wearable band,  called IoT-Q-Band, workflow process. This approach has been deployed in Hong Kong where an electronic wristband linked with a smartphone application in order to track new arrivals at the airports for 14 days \cite{HongKong2020, kupper2011geofencing, normile2020suppress}. Similarly, authorities in the United States have implemented another type of this model using electronic ankle bracelets (ankle monitors) in order to isolate people who refuse to stay in quarantine \cite{Izaguirre2020, Sierra2020}.

\subsection{Drones}
Using drones plays an important role during the quarantine time to decrease the number of COVID-19 cases by lowering the interaction of health care workers with patients and contaminated areas. For example, drones in this phase can assist health care workers and patients by disinfecting areas or delivering medical treatments to patients \cite{lancet2020covid}.

\subsubsection{Disinfectant Drone}
\begin{figure}[h] \centering
    \includegraphics[scale=.38]{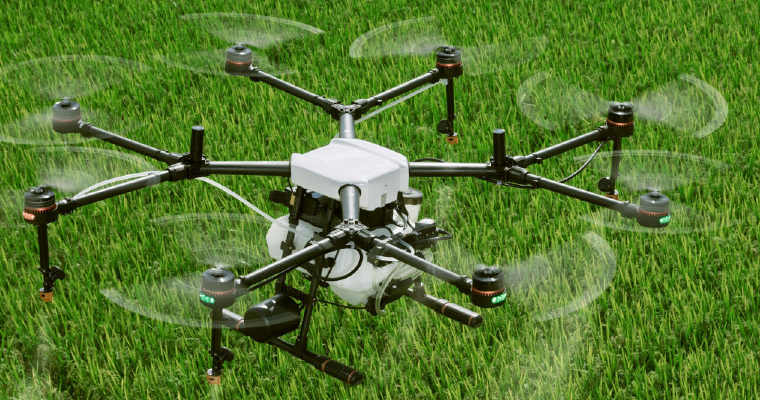}
    \caption{Disinfectant Drone\cite{chamola2020comprehensive}.}
    \label{image:disinfectantDrone}
\end{figure}

Keeping areas sanitized and disinfected during the quarantine period is very important, and this can be achieved by using a particular type of drone, called a Disinfectant Drone\cite{disinfect2020} (See Fig. \ref{image:disinfectantDrone}). These drones can reduce the contamination of the virus and also prevent health care workers from getting infected. DJI company produced this drone with the ability to disinfect one hundred meters in one hour. This type of drone has also been used in Spain for disinfecting purposes \cite{2Drones2020}.

\subsubsection{Medical/Delivery Drone}

\begin{figure}[h] \centering
    \includegraphics[scale=0.35]{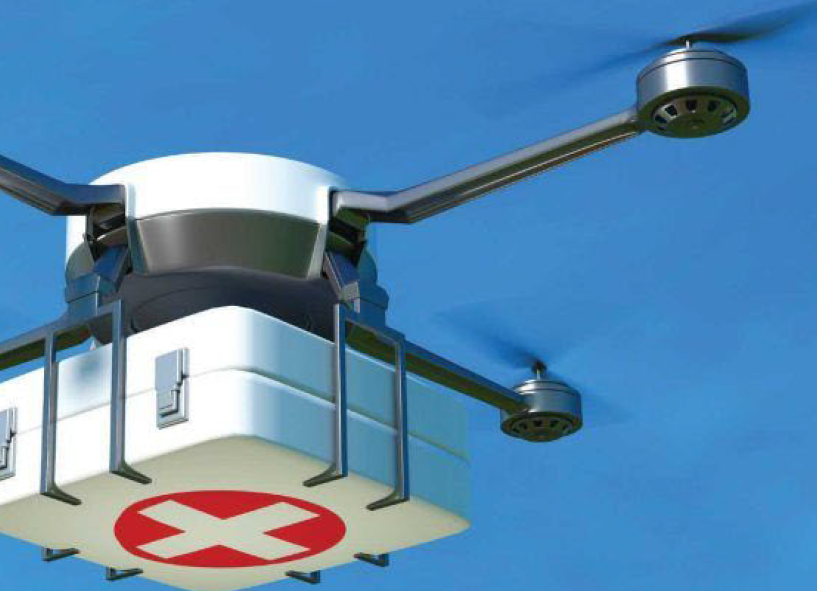}
    \caption{Medical Drone transferring medical related \cite{chamola2020comprehensive}.}
    \label{image:medDrone}
\end{figure}

Medical drones have shown their efficiency at the early stage of \mbox{COVID-19} where they transfer the \mbox{COVID-19} test kits, samples, or medical supplies between labs and medical centers to eliminate human interactions. Additionally, this type of drones usually reduce hospital visits and increase access to medical care by delivering medical treatments to patients or another medical center rapidly. For example, using medical drones in China and Ghana has increased the speed of diagnosis by cutting delivery time \cite{Ghana2020medDrone, chinadrone2020}.
Another type of delivery drone during \mbox{COVID-19} produced by Delivery Drone Canada Inc., which can move COVID-related goods, including test kits and swab tests \cite{DeliveryDrone2020}. 
This type of drone can be also used for other purposes such as postal and grocery services while \mbox{COVID-19} confirmed cases are isolated in their homes during the quarantine time\cite{medicalDrone2020, singla2020drone}. Fig. \ref{image:medDrone} illustrates a picture of this type of drone. 


\subsection{Robots}
During the quarantine time phase, robots play an important role in keeping medical staff away from isolated patients \cite{o2020hospital}. For example, robots can be used in different ways, such as capturing respiratory signs and providing assistance to patients with their treatments or food.

\subsubsection{Telerobots}
\begin{figure}[h] \centering
    \includegraphics[scale=.55]{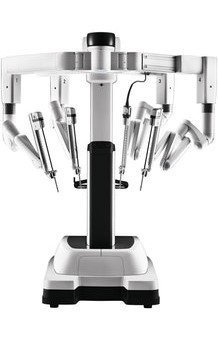}
    \caption{Davinci telerobot can prevent close contact between surgeon and patient during a surgery \cite{daVinci2020}.}
    \label{image:telerobot}
\end{figure}

Telerobots are usually operated remotely by a human and can provide different services such as  remote diagnosis, remote surgeries, and remote treatments for the patients while there is no human interaction during the process \cite{avgousti2016medical}. For example, a nurse can measure patients' temperatures without having interactions with them by using these robots. Another example is daVinci surgical robot, which is operated by a surgeon while the patient is in the safe isolation of plastic sheeting. This helps to prevent infections by performing surgeries remotely \cite{tavakoli2020robotics, yang2020keep}. Fig. \ref{image:telerobot} shows the actual daVinci telerobot. 

\subsubsection{Collaborative Robots}
\begin{figure}[h] \centering
    \includegraphics[scale=.70]{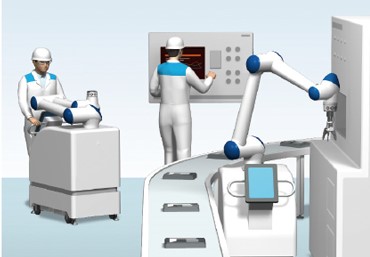}
    \caption{Human Operated Collaborative Robots\cite{collabrobot2013}.}
    \label{image:collabrobot}
\end{figure}

\begin{figure}[h] \centering
    \includegraphics[scale=1.03]{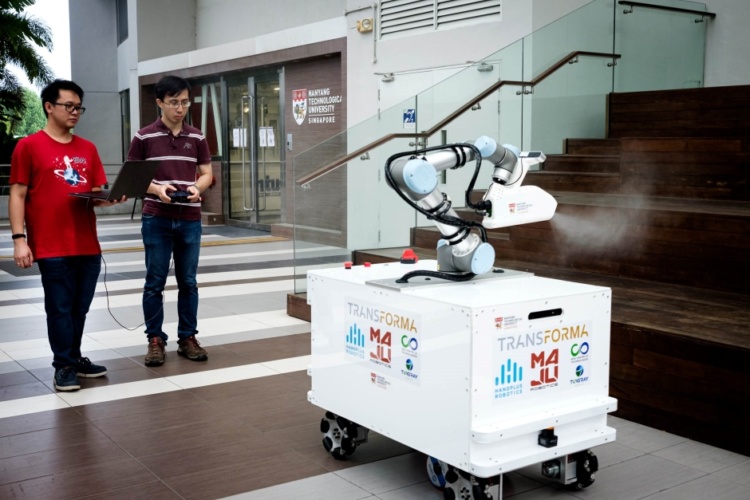}
    \caption{XDBOT Collaborative robot operating by humans for disinfecting contaminated areas\cite{XDBOT2020second}.}
    \label{image:XDBOT}
\end{figure}

Collaborative robots, known as Cobots (Fig. \ref{image:collabrobot}), are recommended robots if there is a need for an operation performed by humans. They are not as beneficial as telerobots for this pandemic, but during a quarantine, this type of robot can lower health care workers' fatigue as well as track their interactions with patients \cite{tavakoli2020robotics}. For instance, Asimov Robotics in India is designed for quarantine time to help patients in isolated areas with tasks such as preparing food and providing medication and also preventing health care workers from being in that area \cite{EconomicTimes2020, cobot2020covid}.  Another example of this robot during this phase is the eXtremeDisinfection robot (XDBOT) (shown in Fig. \ref{image:XDBOT}) which is implemented by Nanyang Technological University in Singapore. This robot can disinfectant hard to access areas, such as under a bed, and also can be wirelessly operated on a mobile platform to avoids any close contact between humans and contaminated areas \cite{cobot2020covid, XDBOT2020second}.

\subsubsection{Autonomous Robots}
\begin{figure}[h] \centering
    \centering
    \includegraphics[scale=.18]{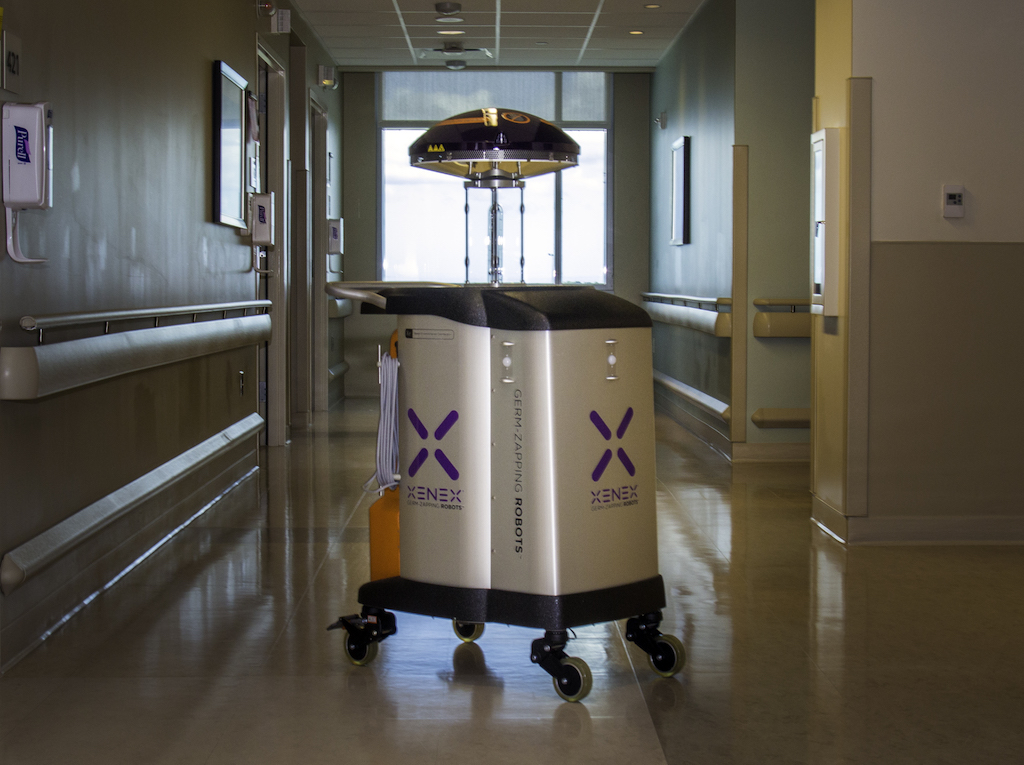}
    \caption{Xenex Disinfecting Autonomous Robots\cite{Xenex2018}.}
    \label{image:Xenex}
\end{figure}

Autonomous robots have been widely used during quarantine time phase. They work with fewer or no human interactions and can be utilized in different scenarios in order to sterilize contaminated areas in hospitals, carry patients' treatments, and check their respiratory signs. These will result in decreasing the risk of infection for the healthcare workers while the patients are isolated in their rooms \cite{tavakoli2020robotics, ackerman2020autonomous}. 
For example, the disinfection robot created by Xenex\cite{Kim2020} is capable of cleaning and disinfecting areas of viruses and bacteria. Fig, \ref{image:Xenex} illustrates how the Xenex robot breaks down the virus using UV lights. Another example is UVD robots developed by a Danish company are used for disinfecting hospitals with their strong UV light, which destroys the DNA of the virus \cite{Danish2020}.

\subsubsection{Social Robots} 
\begin{figure}[h] \centering
    \includegraphics[scale=.58]{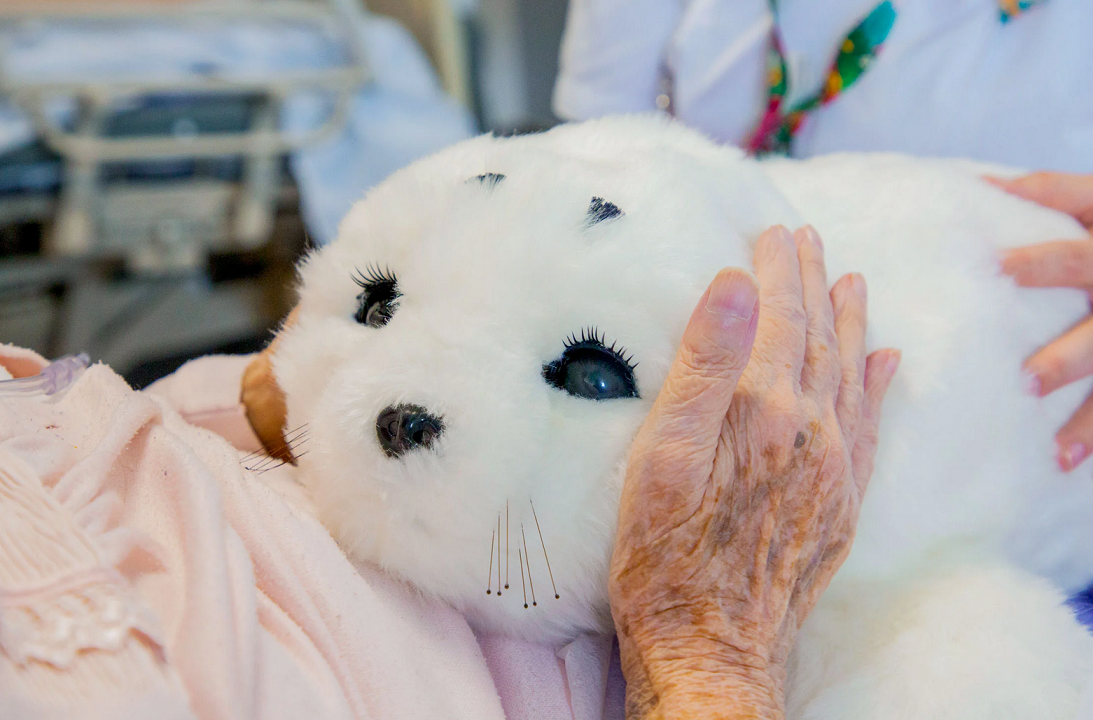}
    \caption{Paro Social Robot can prevent mental effects of quarantine \cite{SocialBot2020}.}
    \label{image:socialbot}
\end{figure}

According to the CDC \cite{CDC2020}, isolating and quarantining patients can potentially cause mental health problems as well. To prevent this, social robots are designed to communicate with patients during that time. The functionality of these robots in this pandemic is to help reduce mental fatigue and strain during a quarantine and period of physical distancing \cite{tavakoli2020robotics}.
One example of these robots is Paro\cite{SocialBot2020}, which can help patients during their isolation as a stress-relief device, as is shown in Fig. \ref{image:socialbot}.

\subsection{IoT Buttons} 
The use of the IoT button in response to the \mbox{COVID-19} pandemic can help to track patients during a quarantine. The Sefucy IoT button\cite{SefucyIoTButton2020} was originally designed for tracking lost or missing children, but with the outbreak of \mbox{COVID-19}, this IoT device has been used for emergency notifications during a quarantine. If the health condition of a confirmed case isolated at home gets worse, by pressing the button, a healthcare provider will be alerted, or family members will be notified in case of an emergency. 
 
 
\subsection{Smartphone Applications}
The most critical part of a quarantine is keeping track of patients while they are isolated. Tracking patients using smartphones during quarantine time is another widely used approach to mitigate and control the spread of this virus.

\subsubsection{Social Monitoring}
In Russia, a mandatory surveillance application called Social Monitoring \cite{Russia2020} has been developed by the government to track patients who are diagnosed with \mbox{COVID-19} and have to be isolated in their homes. Using this approach, authorities can track patients after the application is installed on the patients' smartphones.  Patients are required to ask for QR (Quick Response) code every time they want to leave home or quarantine areas. This code represents their identification to the authorities, which allows them to monitor patients \cite{RussiaPrivacy2020, Meduza2020, kupper2011geofencing}.


\subsubsection{Selfie app}
This application was developed in Poland integrated with Geo-location and facial recognition technology to track patients who have been told to stay at home for 14 days. Patients can reject installing this application, but in return, they will get unexpected visits from authorities. Using the application, patients will be asked to send selfies randomly on a daily basis \cite{SelfieApp2020, geolocation2020, bruce1986understanding}.

\subsubsection{Civitas}
Civitas is a Canadian smartphone application that has been proposed to lower \mbox{COVID-19}'s impact. Using the user's identification code, this application communicates with the authorities to request a permit that allows the user to leave the house. 
 Civitas has the ability to assists suspected cases who need to go outside to buy essential goods in a timely manner. Furthermore, it provides a secure channel that enables physicians to contact patients to monitor their health status \cite{Civitas2020, 2Civitas2020}.

\subsubsection{StayHomeSafe}
StayHomeSafe application is considered as a combination of smartphone applications and wearable devices \cite{HongKong2020}. It has been implemented in Hong Kong where new arrivals at the airports are given a wristband that can be paired with a smartphone in order to set the quarantine location with the advantage of geofencing technology used by the application \cite{kupper2011geofencing}.

\section{Phase III: After Recovery} \label{sec:phase3}
The restrictions put in place response to the COVID-19 pandemic has had a devastating effect on many businesses, marketplaces, and economics. After months of locked down societies and harsh restrictions, nations are gradually and carefully opening up again. This is the phase that everyone needs to experience with extra caution. Social distancing and restrictions on physical services need to be implemented in the way to make sure the virus will not spread again \cite{fong2020nonpharmaceutical}. In this section, we highlight the role of IoT technology in combating the COVID-19 pandemic after lockdown.

\subsection{Wearables}
Since employers are gradually bringing workers back to the workplaces, students are backing to schools, and the economy is bouncing back for reopening, there should be some protection techniques in order to keep everyone safe from this virus. Contact tracing and social distancing are two key points to be considered for safely reopening. Wearables are the devices that can be utilized to trace users' close contacts with other people and also alert them if social distancing is not maintained \cite{Zielinski2020Wearables}.

\subsubsection{EasyBand}
\begin{figure} \centering
    \includegraphics[scale=.69]{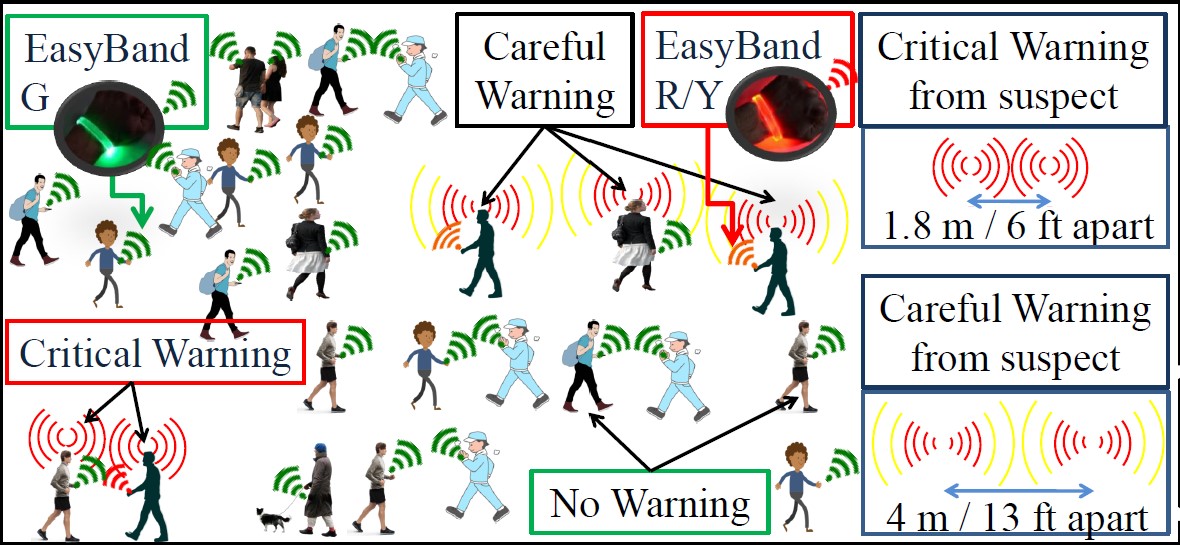}
    \caption{EasyBand process of tracking with its designed rules\cite{tripathy2020easyband}}
    \label{image:easyband}
\end{figure}

\begin{figure}[h] \centering
    \includegraphics[scale=.32]{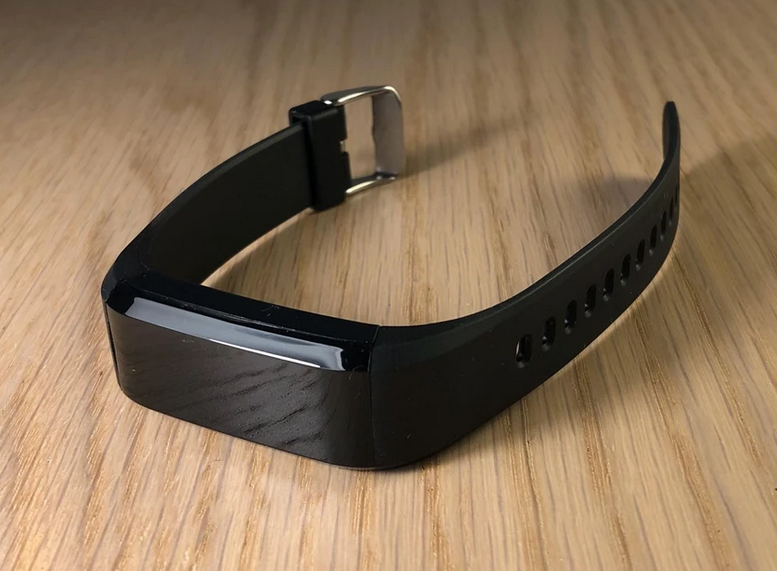}
    \caption{Pact wristband for alerting and tracing\cite{PactWristband2020}.}
    \label{image:pactwristband}
\end{figure}

As countries gradually reopen jobs and marketplaces after the lockdown, EasyBand \cite{tripathy2020easyband} is one of the most effective IoT devices to make sure people are practicing social distancing. This wearable device, which is integrated with the Internet of Medical Things (IoMT), is sensing and capturing data from other devices. EasyBand works within a specific radius and shows potential risk by its LED lights if people are very close to each other. For instance, if someone wearing an Easyband gets close to another person within 4 meters, the band will start beeping to alert both and remind them to keep the distance from each other.
This device has shown better results compared to smartphone apps, and it can be used without any mobile devices. Additionally, it is a cost-effective device that gives people a greater sense of safety and peace of mind. Fig. \ref{image:easyband} represents the workflow of this wearable. A similar example for this device is Pact wristband\cite{PactWristband2020}(see Fig. \ref{image:pactwristband}), which alerts the closeness of people using a vibrator and buzzer.

\subsubsection{Proximity Trace}
\begin{figure}[h] \centering
    \includegraphics[scale=1.35]{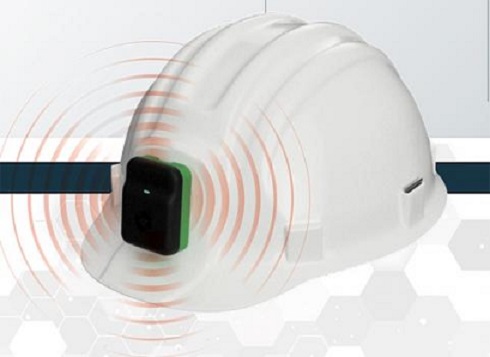}
    \caption{Use of TraceTag on a hard hat\cite{NED2020}.}
    \label{image:proximity}
\end{figure}

\begin{figure}[h] \centering
    \includegraphics[scale=.40]{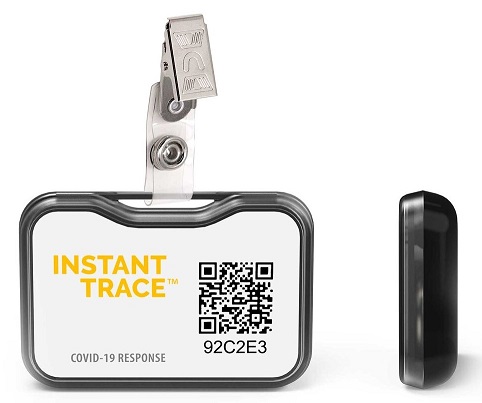}
    \caption{Instant Trace worn as badge \cite{NED2020}.}
    \label{image:instantTrace}
\end{figure}

As industrial workers are coming back to work after the lockdown, there is an essential need for practicing social distancing between them while they work together. Proximity Trace\cite{NED2020} helps industrial workers maintain social distancing in practical ways. This device, which can be attached to a hard hat or body, alerts workers when they get close to each other with a loud sound. Using this device, workers will be able to concentrate on their work without worrying about contamination from the virus. Fig. \ref{image:proximity} shows how this trace can stick to the industrial workers' hard hat. Also, Instant Trace, shown in Fig. \ref{image:instantTrace}, worn as a badge has the same functionality that helps employees to maintain social distancing and trace the infected employee's contacts\cite{instantTrace2020}.

\subsection{Drones}
As the pandemic enters the After Recovery phase, many drones have been used in response to the reopening, which helps businesses continue working in a safe manner. Increasing social awareness by monitoring the crowds and broadcasting information is the main purpose of implementing these devices during this phase \cite{DroneRecovery2020}.

\subsubsection{Surveillance Drone}
\begin{figure}[h] \centering
    \includegraphics[scale=.32]{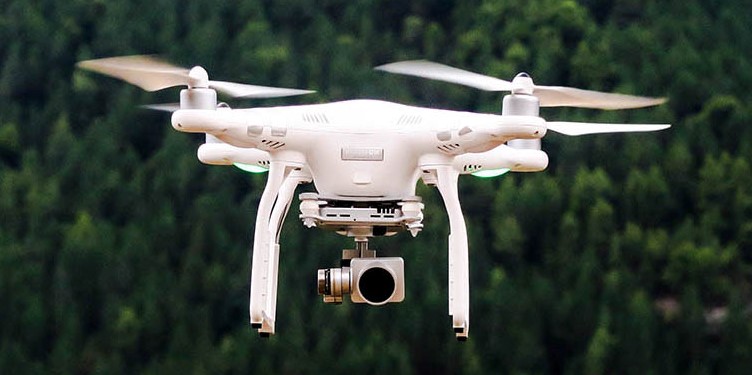}
    \caption{Surveillance Drone\cite{SurvDroneImage2020}.}
    \label{image:surv.Drone}
\end{figure}

Surveillance drone was designed and developed as an effective way to monitor crowds in case of people's failure to do social distancing.  MicroMultiCopter \cite{marr2020robots} made in China and Cyient \cite{Cyient2020} from India are two types of this drone (Fig. \ref{image:surv.Drone}). The MicroMultiCopter drone has also been equipped with speakers to announce important information from the authorities which will be discussed in the next type of drone within this phase.

\subsubsection{Announcement Drone}
\begin{figure}[h] \centering
    \includegraphics[scale=.40]{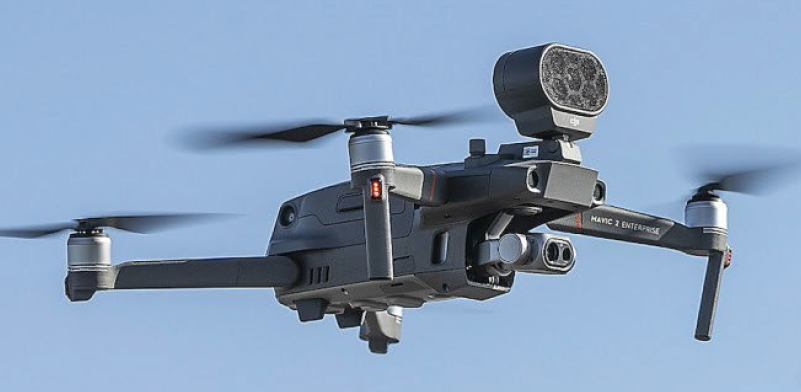}
    \caption{Announcement Drone\cite{chamola2020comprehensive}.}
    \label{image:Announce.Drone}
\end{figure}

This type of drone is mainly designed for broadcasting in areas with low accessibility to the Internet. For example, authorities in Spain and other European countries used this type of drone to announce the practice of social distancing and other guidelines with loudspeakers \cite{1Drones2020, 2Drones2020}. Kuwait is another country that used this drone to broadcast ``go home" messages to people in crowds\cite{singla2020drone} (see Fig.\ref{image:Announce.Drone}).

\subsubsection{Multipurpose Drone}
\begin{figure}[h] \centering
    \includegraphics[scale=.41]{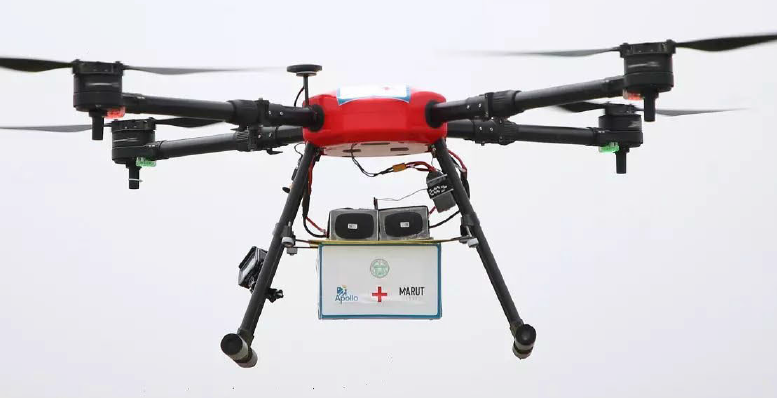}
    \caption{All facilities at once, Multipurpose Drone \cite{chamola2020comprehensive}.}
    \label{image:multiDrone}
\end{figure}

A multipurpose drone, called Corona Combat,\cite{multipurpose2020} has been implemented in China with the combination of all other types of drones that can cover all of the proposed goals mentioned in three phases at once. This drone can be deployed in any \mbox{COVID-19} phases. Fig. \ref{image:multiDrone} shows this drone with all the characteristics from other drones.

\subsection{Robots}
\begin{figure}[h] \centering
    \includegraphics[scale=.22]{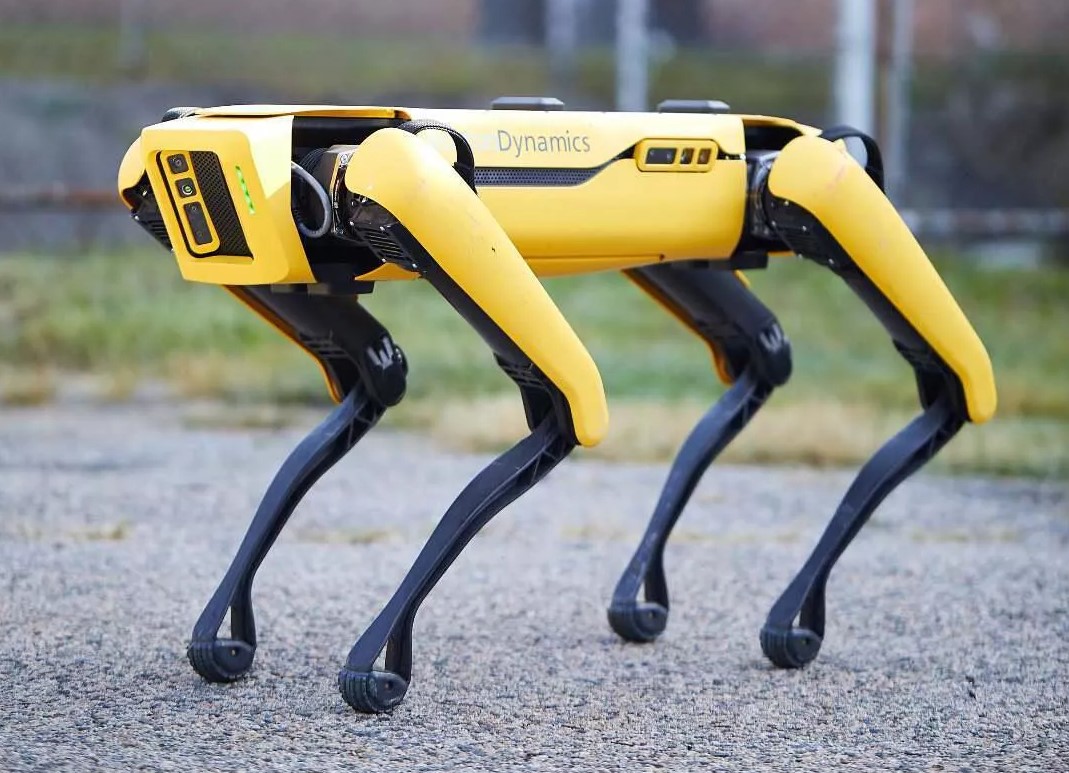}
    \caption{``Spot" Social Distancing Robot\cite{spot2020}.}
    \label{image:spot}
\end{figure}

As schools are opening, businesses are recovering, cars are back on the roads and people are returning to their daily commutes, the COVID-19 pandemic is turning into the next step which is after lockdown or after recovery. In this phase,  everybody needs to know the importance of social distancing everywhere to mitigate the spread of the virus.  
In response to this phase of \mbox{COVID-19}, autonomous robots can be used to control social distancing. For instance, Spot \cite{spot2020covid}, a four-legged robot designed in Singapore to be similar to a dog,  reminds people to practice social distancing in public places. While this robot can be controlled remotely, it is also capable of transferring data to a web interface for further monitoring. \cite{nalewicki2020singapore}. Fig. \ref{image:spot} is the Spot robot for monitoring the practice of social distancing.
\subsection{Smartphone Applications}
The use of IoT in healthcare is now expanding, and the major benefits are cost-effectiveness, efficient monitoring, appropriate treatment, fewer mistakes, and exceptional diagnoses \cite{joyia2017internet, nausheen2018healthcare}. Some smartphone applications have been developed specifically in response to the pandemic’s challenges associated with reopening step which will be covered in this section.

\subsubsection{Aarogya Setu}
Aarogya Setu\cite{1Aarogya2020, 2Aarogya2020} is a  contact tracing application for people to use on their smartphones to increase awareness of and fight against this virus. Aarogya Setu is designed for better communication between health service providers and people. In the application, the user will be asked if he or she has any symptoms of \mbox{COVID-19} or has recently traveled internationally. Analyzing the input data from the users along with their tracking information Aarogya Setu is able to notify the user if he or she has had contact with someone who is already or later becomes a confirmed case.

\subsubsection{TraceTogether}
Singapore launched an application called TraceTogether\cite{TraceTogether2020} to capture data using an encrypted ID from people who were in close contact with each other. The captured data will not be used until a close contact identification is established \cite{Trace2020, SingaporeApp2020}. This data includes the duration of the visit and the social distance will be stored for 21 days for contact tracing purposes in the future. 

\subsubsection{Hamagen}
This contact tracing application was developed in Israel. Hamagen uses GPS technology to find out if the user has been in close contact with a person who tested positive for \mbox{COVID-19}. In this application, for the sake of privacy, private data will not leave the smartphone until the user agrees on \cite{Israel2020}.

\subsubsection{Coalition}
 Coalition\cite{coalition2020} uses IoT technology and blockchain \cite{alladi2019blockchain,8960432,8818320,srivastava2020future} platform to provide a secure  contact tracing approach. In this app, users are assigned with random IDs so that with the detection of any new cases, the users who were in close contact with those cases will be notified.

\subsubsection{BeAware Bahrain}
 BeAware Bahrain is a contact tracing application implemented in Bahrain that alerts people when they are approaching contaminated areas with a detected \mbox{COVID-19} case or if they were in close contact with a confirmed case. Also, this application monitors the location of self-isolated cases for 14 days and allows users to leave quarantine areas for testing appointments, which means this app is also applicable for the second phase of this pandemic \cite{Bahrain2020, Bahrain2020second}.
 
\subsubsection{eRouska}
This application monitors and captures any close proximity between its users. If one of the users' test gets positive for \mbox{COVID-19}, eRouska will notify the others regarding the probable infection so that they can take action about their health situation \cite{erouska2020}.
 

\subsubsection{Social Media - Whatsapp}
As of April 2020, the world has about 3.8 billion users on social media \cite{SocialmediaStat2020}. This number of users creates a great opportunity to implement telemedicine healthcare support using social media applications during this pandemic. One of the most popular applications is Whatsapp. This application provides this chance for patients to consult remotely with their physicians using virtual meetings which will lead to decreasing hospital visits by patients. Using this method is applicable to all phases during the \mbox{COVID-19} pandemic \cite{machado2020social}.

\section{Discussion and Future Work} \label{sec:Discussion}
COVID-19 is  considered as both a global health crisis and an international economic threat.  The restrictions put in place response to the COVID-19 pandemic has had a devastating effect on many businesses, marketplaces, economics, society and our lives. The full health, social, and economical consequences of this pandemic and its restrictions will take time to be fully recognized and quantified, however, there are lots of ongoing efforts in research and industrial communities to utilize different technologies to detect, treat, and trace the virus to mitigate its impacts. Internet of Things (IoT) technology has shown promising results in early detection, quarantine time, and after recovery from COVID-19, however, as we learn more about the virus and its behavior we should adjust and improve our approaches in different phases. For example, it would be interesting to integrate Artificial Intelligence (AI) and IoT technology in order to use AI power to minimize interactions between health care workers and patients in all phases. Another example is using touchless technology with the help of other inputs (such as gesture and voice) will be  efficiently lower the spread of the disease and end the pandemic sooner \cite{agarwal2020unleashing}. Further research needs to be done on convincing  confirmed cases  of \mbox{COVID-19} to remain in quarantine to mitigate the spread of the virus. Moreover, how IoT devices can help isolated patients efficiently for their daily life.  After lockdown, as businesses and marketplaces are opening gradually, how the IoT devices can be incorporated in businesses to cover both safety and efficiency. Answer to those questions will attract considerable attention in both research and industrial disciplines and open new research avenues in this area. 

One of the main concerns about using IoT devices in different phases of this pandemic is privacy issue where patients are asked to share their information. Definitely, it is a big concerns for every patients so defining secure channels for communications or utilizing different encryption techniques before sharing private information would be a possible research area. 

Having IoT-enabled Smart cities can be extremely helpful in combating the current and future pandemic through collaboration between medical centers, cities, etc.\cite{smartcityInsights2020}. Along with aforementioned IoT applications, Allam et al. \cite{allam2020coronavirus} highlights the importance of the concept of Smart City networks while the world is struggling with the \mbox{COVID-19} pandemic. Smart City infrastructure can help people maintain social distancing by the implementation of smart transportation systems  including crowd monitoring, smart parking, and traffic re-routing \cite{gupta2020enabling}. As a part of smart living in the Smart City, smart home IoT-based technologies can also reduce the infection rate of \mbox{COVID-19}. For instance, Smart home doorbells and security systems can be implemented for preventing users from touching surfaces so that there will not be any contamination of the virus by touching those kinds of surfaces \cite{zhang2017security, singh2020internet, smartHome2020}.

\section{Conclusion} \label{sec:Conclusion}
While the world is struggling with \mbox{COVID-19} pandemic, many technologies have been implemented to fight against this disease. One of these technologies is the Internet of Things (IoT), which has been widely used in healthcare industry. During COVID-19 pandemic, this technology has shown very encouraging results dealing with this disease. 
In this paper, we conduct a survey on the recent proposed IoT devices aiming to assist health care workers and authorities during the COVID-19 pandemic. we review the IoT-related technologies and their implementations in three phases including ``Early Diagnosis," ``Quarantine Time," and ``After Recovery."  In each phase, we evaluate the role of IoT enabled/linked technologies including wearables, drones, robots, IoT buttons, and smartphone applications in combating \mbox{COVID-19}. 
IoT technology can be extremely efficient for this pandemic, but it is also critical to consider the privacy of data. By implementing IoT technology properly in a secure way, more patients, with peace of mind, can participate in their treatment journey using IoT devices. As a result, authorities and health care workers can perform a better action regarding the pandemics. Consequently, the impact of these types of disease including the infection, and hospitalization, and death rate can be significantly reduced.

\bibliographystyle{IEEEtran}
\bibliography{mybib}

\end{document}